\documentstyle[epsfig,aps,preprint,prl]{revtex}

\begin{document}

\title{The U(5)--O(6) transition in the Interacting Boson Model and
  the E(5) critical point symmetry}

\author{J.M. Arias$^1$, C.E. Alonso$^1$, A. Vitturi$^2$,
  J.E. Garc\'{\i}a-Ramos$^3$, J. Dukelsky$^4$, and A. Frank$^5$ }

\address{$^1$ Departamento de F\'{\i}sica At\'omica, Molecular y Nuclear,
Facultad de F\'{\i}sica \\ Universidad de Sevilla, Apartado~1065, 
41080 Sevilla, Spain \\
$^2$ Dipartimento di Fisica Galileo Galilei,
Via Marzolo 8 35131 Padova, Italy \\
$^3$ Departamento de F\'{\i}sica Aplicada, Universidad de Huelva, 
21071 Huelva, Spain \\
$^4$ Instituto de Estructura de la Materia,
CSIC, Serrano 123, 28006 Madrid, Spain \\
$^5$ Instituto de Ciencias Nucleares,
UNAM, Circuito Exterior C.U., M\'exico D.F., 04510 M\'exico}

\maketitle

\begin{abstract}
The relation of the recently proposed E(5) critical point
symmetry with the interacting boson model is investigated. 
The large-N limit of the interacting boson model at the critical
point in the transition from U(5) to O(6) is obtained by
solving the Richardson equations. It is shown explicitly
that this algebraic calculation leads to the same results as the
solution of the Bohr differential equation with a $\beta^4$ potential.
\end{abstract}

\vspace{2cm}


\newpage

The study of phase transitions is one of the most exciting topics in
Physics. Recently the concept of critical point symmetry has been proposed by
Iachello\cite{E5}. These kind of symmetries apply when a quantal system
undergoes transitions between traditional dynamical symmetries.
In Ref.\cite{E5} the particular case of the Bohr
Hamiltonian\cite{BMII} in Nuclear Physics was worked out.
In this case, in the situation in which the potential energy surface
in the $\beta$-$\gamma$ plane
is $\gamma$-independent and the dependence in the $\beta$ degree of
freedom can be modeled by an infinite square well, 
the so called E(5) symmetry appears. This situation is expected to be
realized in actual nuclei when they undergo a transition from
spherical to $\gamma$-unstable deformed shapes. The E(5)
symmetry is obtained 
within the formalism based on the Bohr hamiltonian, but 
it has also been used in connection with the Interacting
Boson Model (IBM)\cite{IBM}. Although this is not the form it was
originally proposed\cite{E5}, it has been in fact argued that moving
from the spherical to the $\gamma$-unstable deformed case within the IBM
one should reobtain, at the critical point in the transition,
the predictions of the E(5)
symmetry.  This correspondence is supposed to be valid in the limit of
large number N  of bosons, but the
calculations with the IBM should provide predictions for
finite N as stated in Ref.\cite{E5exp}. 
In this letter, on one hand we calculate exactly the large N limit of
the IBM at the critical point in the transition from U(5) (spherical case) 
to O(6) (deformed $\gamma$-unstable case). On the other hand, we solve
the Bohr differential equation for a $\beta^4$ potential. Both
calculations lead to the same results and are not close to those
obtained by solving the Bohr
equation for an infinite square well (E(5) symmetry).   
We also show with two schematic examples that the corrections 
arising from the finite number of bosons 
are important. With this in mind, the IBM calculations still provide a
tool for  including corrections due to the finite number of bosons.


In Ref. \cite{E5} the Bohr Hamiltonian is considered for the case of a
$\gamma$ independent potential, described by an infinite square well
in the $\beta$  
variable. In that case, the hamiltonian is separable in both variables
and if we set
\begin{equation}
\Psi(\beta,\gamma,\theta_i)=f(\beta) \Phi(\gamma, \theta_i) ~
\end{equation}
where $\theta_i$ stands for the three Euler angles, 
the Schr\"odinger equation can be split in two equations. 
The solutions of the ($\gamma,\theta_i$) part were studied in
Ref.\cite{WJ56} and tabulated in Ref.\cite{Bes}. Iachello solved the
$\beta$ part and found that the $f(\beta)$ functions are related to
Bessel functions. The main results are illustrated in Table I and Fig. 1
of Ref.\cite{E5}. These results are obtained from a geometrical
picture and we would like to investigate its relation with the
interacting boson model.

The geometrical
interpretation of the abstract IBM hamiltonian  can be
obtained by introducing a coherent state\cite{GK80,DSI80,BM80} which
allows to associate to it a geometrical shape in terms of the deformation
variables ($\beta$, $\gamma$). 
The basic idea of this formalism is to consider that the
pure quadrupole states are 
globally described by a boson condensate of the form
\begin{equation}
|g; N, \beta, \gamma \rangle = {1 \over \sqrt{N!}} (\Gamma_g^\dag) ^N |0 \rangle ~~,
\label{ground1}
\end{equation}
where the basic boson is given by
\begin{equation}
\Gamma_g^\dagger={1 \over \sqrt{1+ \beta^2}} 
\left[ s^\dagger + \beta \cos \gamma d^\dagger_0 + {1 \over \sqrt{2}} 
\beta \sin\gamma (d^\dagger_2 + d^\dagger_{-2})\right]~~,
\label{ground2}
\end{equation}
which depends on the $\beta$ and $\gamma$ shape variables. The energy
surface is defined as
\begin{equation}
E_N(\beta, \gamma)= \langle g; N, \beta, \gamma| \hat H | g; N, \beta,
\gamma \rangle ~~,
\label{EVH}
\end{equation}
where $\hat H$ is the IBM hamiltonian. Here we are interested in the case
in which the hamiltonian undergoes a transition 
from U(5) to O(6) and, consequently, the
corresponding potential energy surfaces are $\gamma$-independent.

In order to investigate the geometrical limit of the IBM in the
transitional class going from U(5) (spherical) to O(6) (deformed
$\gamma$-unstable) the most general (up to two-body terms) IBM
hamiltonian is,
\begin{equation}
\label{ham1}
\hat H =
\varepsilon_d \hat n_d +
\kappa_0 \hat P^\dag \hat P
+\kappa_1 \hat L\cdot \hat L+
\kappa_2 \hat Q^{\chi=0} \cdot \hat Q^{\chi=0} +
\kappa_3 \hat T_3\cdot\hat T_3 +\kappa_4 \hat T_4\cdot\hat T_4
\end{equation}  
where $\hat n_d$ is the $d$ boson number operator, and 
\begin{eqnarray}
\label{P}
\hat P^\dag&=&\frac{1}{2} ~ (d^\dag \cdot d^\dag - s^\dag \cdot s^\dag), \\
\label{L}
\hat L&=&\sqrt{10}(d^\dag\times\tilde{d})^{(1)},\\
\label{Q}
\hat Q^{\chi=0}&=& (s^{\dagger}\times\tilde d
+d^\dagger\times\tilde s)^{(2)} ,\\
\label{t3}
\hat T_3&=&(d^\dag\times\tilde{d})^{(3)},\\
\label{t4}
\hat T_4&=&(d^\dag\times\tilde{d})^{(4)}.
\end{eqnarray} 
The scalar product is defined as 
$\hat T_L\cdot \hat T_L=\sum_M (-1)^M \hat T_{LM}\hat T_{L-M}$, where 
$\hat T_{LM}$ corresponds to the $M$ component of the operator 
$\hat T_{L}$. The operators $\tilde d_{m}=(-1)^{m} d_{-m}$ and $\tilde
s=s$ are introduced to ensure the correct tensorial character under
spatial rotations. 
The corresponding energy surface is obtained from Eq. (\ref{EVH})

\begin{eqnarray}
E(N,\beta)  & = &  \frac{N}{1+\beta^2} \left[ 5 \kappa_2 + \beta^2
  (\varepsilon_d+6 \kappa_1+\kappa_2+\frac{7}{5}
  \kappa_3+\frac{9}{5}\kappa_4) \right] \nonumber \\
   & + &  \frac{N (N-1)}{(1+\beta^2)^2}\left[
  \frac{(1-\beta^2)^2}{4} \kappa_0 + 4 \beta^2 \kappa_2 +
  \frac{18}{35} \beta^4 \kappa_4 \right].
\label{supgen}
\end{eqnarray}

The condition to find the critical point is

\begin{equation}
\left(d^2 E(N,\beta)/d\beta^2\right)_{\beta=0}=0
\label{loc-crit}
\end{equation}
and gives the following relation among the hamiltonian parameters
\begin{equation}
\varepsilon_d=-6 \kappa_1+ 4 \kappa_2-\frac{7}{5}
  \kappa_3-\frac{9}{5}\kappa_4 + (N-1) (\kappa_0-4 \kappa_2).
\label{crit}
\end{equation}
Thus the most general energy surface at the critical point in the
U(5)--O(6) phase transition is
\begin{equation}
E^{crit}(N,\beta) =  5 N \kappa_2 + N (N-1)\left[\frac{\kappa_0}{4} +
  \left(\kappa_0 - 4 \kappa_2 + \frac{18}{35} \kappa_4\right)
  \frac{\beta^4}{(1+\beta^2)^2} \right]. 
\label{supencrit}
\end{equation}
These expressions are consistent with those obtained in
Ref. \cite{Alex} for a slightly different hamiltonian.
Note that (\ref{supencrit}) completely defines the form of the potential up 
to a scale and an energy translation.
The expansion of this critical energy surface around $\beta=0$ is 

\begin{equation}
E^{crit}(N,\beta) \approx 5 \kappa_2 N + \frac{\kappa_0}{4} N (N-1)+
N (N-1) \left(\kappa_0 - 4 \kappa_2 + \frac{18}{35} \kappa_4\right) 
\left[ \beta^4 - 2 \beta^6 +  \dots \right].
\end{equation}
whose leading term is $\beta^4$. Alternatively, one can carry out the
transformation $\beta^2 /(1+\beta^2) \rightarrow \bar \beta^2$ and
finds $\bar \beta^4$ as the critical potential.

\begin{figure}
\begin{center}
\mbox{\epsfig{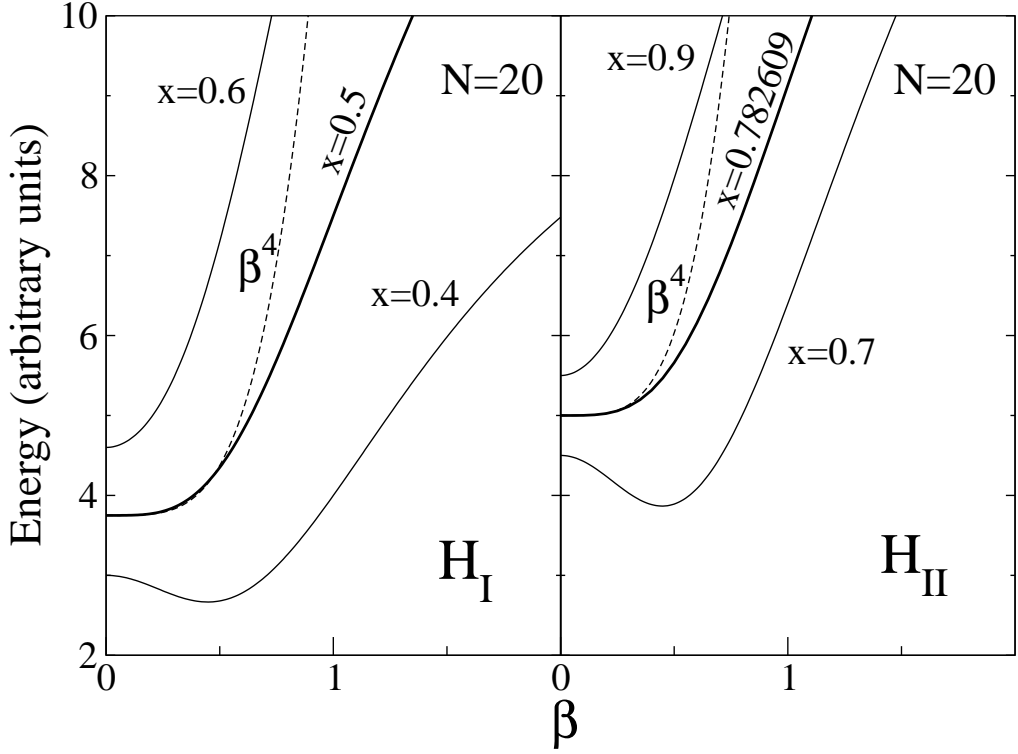}}
\end{center}
\caption{Representation of the energy surfaces for $N=20$ as functions of the
  shape parameter $\beta$ obtained for two
  schematic hamiltonians, Eq. (\ref{HPP}) (left panel) and Eq. (\ref{HQQ})
  (right panel). In each case three values of the order parameter are
  presented, one at the critical value, one above and one below that
  value. The curves have been arbitrarily displaced in energy so as to
  show clearly the behavior.}
\label{fig1}
\end{figure}

In order to make some calculations to illustrate the large N limit in
the IBM at the critical point in the U(5)--O(6) phase transition and the
corresponding finite N corrections,  
we propose two schematic transitional hamiltonians.  The first one is
\begin{equation}
\hat H_I = x \hat n_d + \frac{1-x}{N-1} \hat P^\dag \hat P ~.
\label{HPP}
\end{equation}
The corresponding energy surface is obtained from Eq. (\ref{supgen})
with $\varepsilon_d=x$, $\kappa_0=\frac{1-x}{N-1}$ and all the rest
of the parameters equal to 0,
\begin{equation}
E_I(N,\beta) = N \left[ x \frac{\beta^2}{1+\beta^2} + \frac{1-x}{4}~
  \left( \frac{1-\beta^2}{1+\beta^2} \right)^2 \right].
\end{equation}
The condition to localize the critical point, Eq. (\ref{crit}), gives
in this case $x_c^I=0.5$. 
In Fig. \ref{fig1} we represent as an example the energy surfaces for
the hamiltonian (\ref{HPP}) (left panel) with three selections for the order
parameter $x$: one at the critical point, one above that value and one
below it. For $x>x_c$ an equilibrium spherical shape is obtained,
while for $x<x_c$ the equilibrium shape is deformed. The value $x_c$
gives a flat $\beta^4$ surface close to $\beta=0$. 

The second schematic hamiltonian we propose is 
\begin{equation}
\hat H_{II} = x \hat n_d - \frac{1-x}{N} \hat Q^{\chi=0} \cdot \hat
Q^{\chi=0} ~, 
\label{HQQ}
\end{equation}
The corresponding energy surface is obtained from
 Eq. (\ref{supgen}) with
 $\varepsilon_d=x$, $\kappa_2=- \frac{1-x}{N}$ and all the rest
of the parameters equal to 0,
\begin{equation}
E_{II}(N,\beta) = - (5+\beta^2) ~\frac{1-x}{1+\beta^2} + N~ x~
  \frac{\beta^2}{1+\beta^2} - 4
  (N-1)(1-x)~\frac{\beta^2}{(1+\beta^2)^2} ~. 
\end{equation}
Condition (\ref{crit}) gives in this case the critical point 
$x_c^{II}=\frac{4N-8}{5N-8}$ that in the large N limit gives 4/5. 

In Fig. \ref{fig1} the corresponding energy surfaces are plotted in
the right panel. Same comments as in the preceding case are in order.
Thus, we conclude that, in the transition from spherical systems to
$\gamma$-unstable deformed ones, the critical point in IBM
should be associated to a  
$\beta^4$ potential rather that to an infinite square well. The question
is then how different are the E(5) predictions from those obtained
with a $\beta^4$ potential?
In order to investigate this point we have solved numerically the Bohr
hamiltonian for a potential $\beta^4$. The results for energies are
presented in Table \ref{tab2} and in Fig. \ref{fig3}. Here we keep the
label $\xi$ used in the E(5) case. It is related
to the label $n_\beta=\frac{n_d-\tau}{2}$, sometimes used in the U(5)
classification, by $n_\beta=\xi-1$, where $n_d$ is the U(5) label and $\tau$
is the O(5) label. 
Particularly interesting are the energy ratios given in Table \ref{tab3}
which have been used in recent
works to identify possible nuclei as critical. In this table the E(5)
and $\beta^4$ values are shown for comparison. The labeling of the
states is $L_{\xi,\tau}$. 

Besides the excitation energies, B(E2) transition probabilities can be
calculated using the quadrupole operator 
\begin{equation}
T^{(E2)}_\mu = t ~\beta ~\left[{\cal D}_{\mu 0}^{(2)}(\theta_i) \cos
  \gamma + \frac{1}{\sqrt{2}} \left({\cal D}_{\mu 2}^{(2)}(\theta_i)
  + {\cal D}_{\mu -2}^{(2)}(\theta_i)\right) \sin \gamma \right] ~~,
\label{E2Bohr}
\end{equation}
where $t$ is a scale factor. In Table \ref{tab3} two important B(E2) ratios
are given for E(5) and $\beta^4$ cases. In Fig. \ref{fig3} the
B(E2) values for a $\beta^4$ potential are shown besides the arrows. They are 
given normalized to the $B(E2;2_{1,1}
\rightarrow 0_{1,0})$ value which is taken as 100. 

Comparing Figs. 1
and Table I in Ref.\cite{E5} with
the present Fig. \ref{fig3} and Table \ref{tab2} we can observe
important differences between E(5) and $\beta^4$ potentials. 
In order to see which is the actual large N
limit of IBM we have performed calculations with the IBM codes for hamiltonians
$H_I$ (Eq. \ref{HPP}) and $H_{II}$ (Eq. \ref{HQQ}) at the critical
point for different number of bosons. These codes allow to
manage a small number of bosons, typically 20. In Fig. \ref{fig4} the
results of these calculations are shown with a full line for
Eq. (\ref{HPP}) and with a dashed line for Eq. (\ref{HQQ}). The values for
E(5) and $\beta^4$ potentials are shown as dotted lines as references. The last
two panels labeled with $R_1$ and $R_2$ refer to the B(E2) ratios
presented in Table \ref{tab3}. 

From Fig. \ref{fig4} it is clear that the finite N effects are
important and depend on the precise form of the hamiltonian
used. However, it is difficult to conclude whether E(5) or
$\beta^4$ is the large N limit of the corresponding IBM
hamiltonian. It is necessary to perform calculations with larger values
of N. Fortunately, Dukelsky et
al.\cite{Duk01} have recovered an exactly solvable model for
pairing proposed by Richardson in the 60's\cite{Richard}. 
Following Ref.\cite{Duk01} we have solved the Richardson's 
equations and obtained the exact eigenvalues 
for the hamiltonians (\ref{HPP}) and (\ref{HQQ})
up to $N=1000$, so approaching the large N limit of the corresponding IBM
hamiltonians. Details of this method will be given in a longer
publication. In Fig. \ref{fig5} we present the
results of these calculations for energy ratios up to $N=1000$
and B(E2) ratios up to $N=40$ together
with the corresponding values for the E(5) symmetry and the $\beta^4$
potential. From this figure
it clearly emerges that the large N limit for the studied IBM hamiltonians 
corresponds to the $\beta^4$ potential. Both 
hamiltonians Eq. (\ref{HPP}) and Eq. (\ref{HQQ})
converge to the same results in the large N limit, although the
corresponding corrections for finite N are quite different (see
Fig. \ref{fig4}).

We conclude that the large N limit of the IBM
hamiltonian at the critical point in the transition from U(5)
(spherical) to O(6) (deformed $\gamma$-unstable) is represented
in the geometrical model  by a $\beta^4$ potential. The results are
similar but not close to those of 
an infinite square well as in the E(5) critical point symmetry. 
The analysis of the IBM energy surface followed by an IBM
calculation, as presented in Ref.\cite{E5Ru}, can provide the
appropriate finite N corrections and thus lead to the 
identification of nuclei at the critical points. In that work a systematic
study of the properties of the Ru isotopes allowed to select the
appropriate form of the hamiltonian. Once it is fixed the construction
of the energy surfaces identify the critical nucleus ($^{104}$Ru in
that case). The corresponding IBM calculation for the critical nucleus
then provides the correct finite N corrections. We believe that this
is a fundamental step if we wish to robustly identify the
spectroscopic properties that signal the presence of criticality in
the atomic nucleus.

\section*{Acknowledgements}

This work was supported in part 
by the Spanish DGICYT under project number BFM2002-03315, by 
CONACyT (M\'exico) and by an INFN-DGICYT agreement.

\begin{table}
\caption{Excitation energies for a $\beta^4$ potential relative to
  the energy of the first excited state.}
\begin{center}
\begin{tabular}{|c|cccc|} 
 & $\xi=1$ & $\xi=2$  & $\xi=3$ & $\xi=4$  \\
\tableline 
$\tau=0$  & 0.00 & 2.39 & 5.15 & 8.20  \\
$\tau=1$  & 1.00 & 3.63 & 6.56 & 9.75  \\
$\tau=2$  & 2.09 & 4.92 & 8.01 & 11.34 \\
$\tau=3$  & 3.27 & 6.26 & 9.50 & 12.95  \\
\end{tabular}
\end{center}
\label{tab2}
\end{table}

\begin{table}
\caption{Energy and B(E2) transition rate ratios 
in the E(5) symmetry and for the $\beta^4$ potential.}
\begin{center}
\begin{tabular}{|c|cccc|cc|} 
 & $E_{4_{1,2}}/E_{2_{1,1}}$ &  $E_{0_{2,0}}/E_{2_{1,1}}$ &
     $E_{0_{1,3}}/E_{2_{1,1}}$  & $E_{0_{2,0}}/E_{0_{1,3}}$ &
     $R_1=\frac{B(E2;4_{1,2} \rightarrow 2_{1,1})}{B(E2;2_{1,1}
     \rightarrow 
     0_{1,0})}$ &  $R_2=\frac{B(E2;0_{2,0} \rightarrow 
     2_{1,1})}{B(E2;2_{1,1} \rightarrow 0_{1,0})}$ \\
\tableline 
E(5)  & 2.20 & 3.03 & 3.59 & 0.84  & 1.68 & 0.86  \\
$\beta^4$  & 2.09 & 2.39 & 3.27 & 0.73 & 1.82 & 1.41  \\
\end{tabular}
\end{center}
\label{tab3}
\end{table}

\begin{figure}
\begin{center}
\mbox{\epsfig{file=espectrob4.eps,height=10cm,angle=0}}
\end{center}
\caption{Schematic spectrum for a $\beta^4$ potential. 
Numbers close to the arrows are B(E2) values. These 
are relative to the transition
  $2_{1,1} \rightarrow 0_{1,0}$ whose B(E2) value is taken as 100.}
\label{fig3}
\end{figure}

\newpage

\begin{figure}
\begin{center}
\mbox{\epsfig{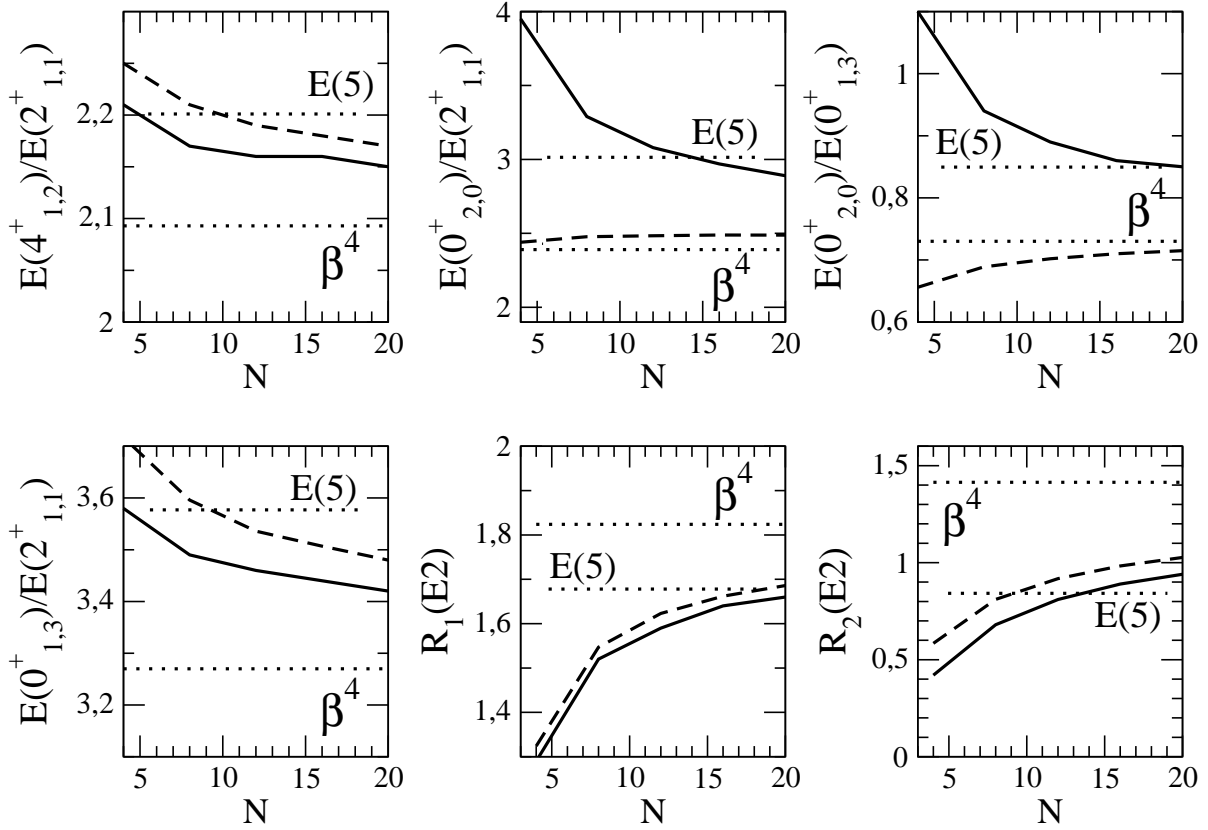}}
\end{center}
\caption{Variation with the number of bosons (up to $N=20$) of
  selected energy and B(E2) ratios for IBM calculations 
  performed at the critical points of hamiltonian (\ref{HPP}) (full
  line) and (\ref{HQQ}) (dashed line). The corresponding E(5) and
  $\beta^4$ values are marked with dotted lines. }
\label{fig4}
\end{figure}

\begin{figure}
\begin{center}
\mbox{\epsfig{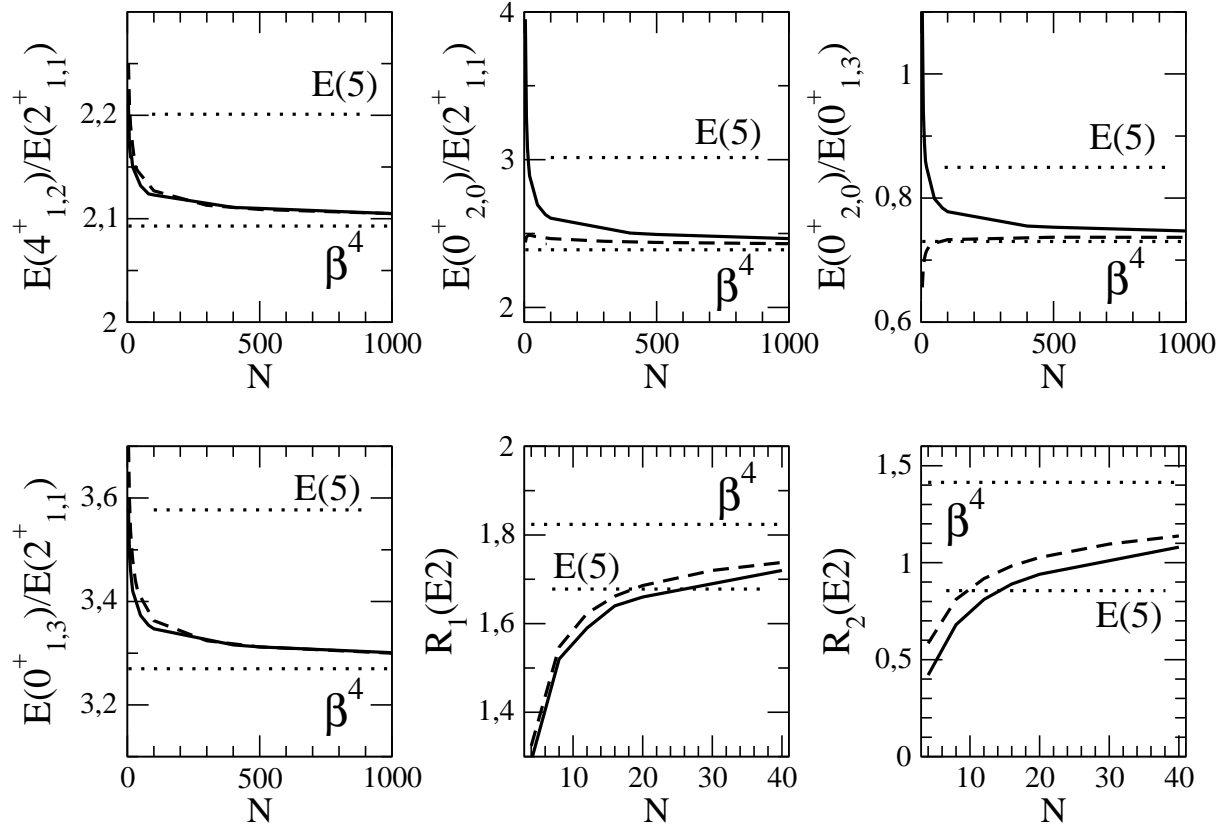}}
\end{center}
\caption{Same as Fig \ref{fig4} but here the number of bosons runs up
  to 1000 in the energy ratios and up to 40 in the B(E2) ratios. }
\label{fig5}
\end{figure}

\end{document}